\documentclass[]{aastex631}

\newcommand{\msol}{\text{M}_\odot}
\newcommand{\mass}{\text{M}}

\newcommand{\rms}[1]{\ensuremath{_{\text{#1}}}}

\newcommand{\atom}[2]{$^{#2}\text{#1}$}
\newcommand{\al}{\atom{Al}{26}}
\newcommand{\fe}{\atom{Fe}{60}}
\newcommand{\seba}{\texttt{SeBa}}

\newcommand{\bhtree}{\texttt{BHTree}}
\newcommand{\amuse}{\texttt{AMUSE}}
\usepackage{savesym}
\savesymbol{tablenum}
\usepackage{siunitx}
\restoresymbol{SIX}{tablenum}

\begin{document}

\title{Towards a unified injection model of short-lived radioisotopes in $N$-body simulations of star-forming regions}

\author[0000-0002-5160-8871]{Joseph W. Eatson}
\affiliation{Department of Physics and Astronomy \\
The University of Sheffield \\
Hicks Building, Hounsfield Road \\
Sheffield, S3 7RH, UK}

\author[0000-0002-1474-7848]{Richard J. Parker}
\altaffiliation{Royal Society Dorothy Hodgkin fellow}
\affiliation{Department of Physics and Astronomy \\
The University of Sheffield \\
Hicks Building, Hounsfield Road \\
Sheffield, S3 7RH, UK}

\author[0000-0002-3286-7683]{Tim Lichtenberg}
\affiliation{Kapteyn Astronomical Institute \\ 
University of Groningen\\
P.O. Box 800, 9700 AV Groningen, NL}

\begin{abstract}

Recent research provides compelling evidence that the decay of short-lived radioisotopes (SLRs), such as \al{}, provided the bulk of energy for heating and desiccation of volatile-rich planetesimals in the early Solar System. However, it remains unclear whether the early Solar System was highly enriched relative to other planetary systems with similar formation characteristics.
While the Solar System possesses an elevated level of SLR enrichment compared to the interstellar medium, determining SLR enrichment of individual protoplanetary disks observationally has not been performed and is markedly more difficult.
We use $N$-body simulations to estimate enrichment of SLRs in star-forming regions through two likely important SLR sources, stellar winds from massive stars and supernovae.
We vary the number of stars and the radii of the star-forming regions and implement two models of stellar wind SLR propagation for the radioisotopes \al{} and \fe{}.
We find that for \al{} enrichment the Solar System is at the upper end of the expected distribution, while for the more supernovae dependent isotope \fe{} we find that the Solar System is comparatively very highly enriched.
Furthermore, combined with our previous research, these results suggest that the statistical role of \al{}-driven desiccation on exoplanet bulk composition may be underestimated in typical interpretations of the low-mass exoplanet census, and that \fe{} is even less influential as a source of heating than previously assumed.

\end{abstract}


\section{Introduction}

\al{} and \fe{}, as well as other Short-Lived Radioisotopes (SLRs) have a significant impact on the formation and evolution of planets, in particular through their influence on planetesimal and protoplanet evolution \citep{lichtenbergWaterBudgetDichotomy2019,2023ASPC..534..907L}.
As these isotopes have a half-life commensurate with planetary formation timescales \citep[0.717\,Myr for \al{}, and 2.6\,Myr for \fe{}, which compares to observations of (proto)planets around 1\,Myr old stars,][]{2020ApJ...904L...6A,2020Natur.586..228S}, they are present throughout the early planetesimal formation process, concentrating onto planetesimals and debris as the disk agglomerates into a protoplanetary system \citep{2023ASPC..534.1031K}.
SLRs are important for planetary formation as they provide a significant source of heating through radioactive decay for nascent protoplanets and planetesimals.
This can increase the rate of chemical segregation within a forming protoplanet \citep{doddsThermalEvolutionPlanetesimals2021}, and is instrumental in the evaporation and removal of volatiles from planetesimals through outgassing \citep{monteuxWaterInteriorStructure2018a,2023Natur.615..854N,2024NatAs...8..290G}.
Volatile outgassing is of particular interest, as removal of H$_2$O and other atmospheric volatiles from planetesimals prior to protoplanetary formation can significantly impact the formation of extrasolar ocean worlds, and instead lead to relatively volatile-poor rocky worlds instead \citep{2021ApJ...913L..20L,2022ApJ...938L...3L}.

It has been found that the Solar System has a significantly higher fraction of both \al{} and \fe{} than the interstellar medium (ISM) \citep{kita26Al26MgIsotopeSystematics2013}, with an isotopic ratio  \atom{Al}{26}$/$\atom{Al}{27} ($Z\rms{26Al})$ of $\approx \num{5e-5}$ based on observations of decay products in chondritic meteorites \citep{2008E&PSL.272..353J,2013M&PS...48.1383K}. The measurement of \fe
{} is much more controversial, though recent estimates \citep[e.g.][]{2021ApJ...917...59C} suggest an isotopic ratio \atom{Fe}{60}$/$\atom{Fe}{56} ($Z\rms{60Fe})$ of $\approx \num{1e-6}$, which is also significantly higher than estimates of the abundance of this isotope in the ISM.

Some researchers have suggested that the parent giant molecular cloud (GMC) of the Solar System had a higher level of SLR enrichment compared to the ISM, due to sequential star formation \citep{2012A&A...545A...4G}, whereas other authors have proposed either internal production or delivery from an external source.
Although \al{} can be produced internally through cosmic ray spallation, this is insufficient and does not reflect the isotropic spatial distribution we observe in the SLR distribution in the Solar System nor the fractionation between spallation and nucleosynthetic processes \citep{lugaroRadioactiveNucleiCosmochronology2018,parkerBirthEnvironmentPlanetary2020}. \fe{} cannot be produced via cosmic ray spallation \citep{lugaroRadioactiveNucleiCosmochronology2018}.

Instead, one of the main proposed external delivery methods is through stellar winds, in which significant quantities of \al{} are produced in the cores of massive stars \citep{palaciosNewEstimatesContribution2005,2009ApJ...696.1854G}, the convective core circulates \al{} to the surrounding envelope, and is expelled as a part of the stellar wind.
Whilst some \fe{} can be emitted through massive stellar winds, this quantity is many orders of magnitude lower than the total wind mass of \al{} \citep{limongiPresupernovaEvolutionExplosive2018,brinkman26AluminumMassiveBinary2021}.
Additionally, large quantities of \al{} and \fe{} are also deposited through supernovae; however massive stars take upwards of \SI{10}{Myr} to explode \citep{limongiPresupernovaEvolutionExplosive2018}, by which time disks are likely already far into producing protoplanets, preventing efficient isotope transfer from the outflow to the planetary system before planetesimal formation \citep{2023ASPC..534..717D,2023ASPC..534..539M}.
Finally, there is additional evidence to suggest that a third category of enrichment is potentially important, which is the case of the winds of ``interloping'' asymptotic giant branch (AGB) stars depositing SLRs into a recently formed star-forming region \citep{karakasStellarYieldsMetalRich2016,wasserburgIntermediatemassAsymptoticGiant2017}.
AGB stars wandering into star-forming regions have been observed \citep{parkerIsotopicEnrichmentPlanetary2023}, however there has been limited study on the impact of such interlopers, or the probability of this occurring with a star-forming region.

In this paper, we report the results of a series of $N$-body simulations of star-forming regions using \amuse{} \citep{pelupessyAstrophysicalMultipurposeSoftware2013} where massive stellar winds and supernovae inject \al{} and \fe{} into protoplanetary disks surrounding low-mass stars.
The region size, population and density are varied significantly for each sub-set of simulations to explore the star-forming region parameter space.
This model utilises wind and supernova mass loss rates of SLRs from work by \citet{limongiPresupernovaEvolutionExplosive2018}, as well as statistical models of star-forming regions and stellar populations.
Our model aims to be more comprehensive than previous attempts to simulate SLR injection within star-forming regions \citep{parkerShortlivedRadioisotopeEnrichment2023,nicholsonSupernovaEnrichmentPlanetary2017,zwartFormationSolarsystemAnalogs2019} by including all facets of massive star-based SLR injection, though there are a number of additional features and avenues of research that we will discuss in brief at the end of the paper. 
The following section details the methodology of our work, in particular the programming framework and implementation of various simulation features, Section \ref{sec:results} details our results, and Section \ref{sec:conclusion} concludes and discusses further research into this particular topic.  

\section{Methodology}
\label{sec:methodology}

The simulation code used throughout this paper is written in Python, using the \amuse{} framework to incorporate pre-existing stellar evolution and $N$-body dynamics code.
At the beginning of every time step, the code calculates positions and velocities for each star in the simulated region, as well as the individual star properties such as mass loss rates.
The stellar properties are calculated using the \seba{} stellar evolution code \citep{portegieszwartPopulationSynthesisHighmass1996,toonenSupernovaTypeIa2012}, while the dynamical properties are calculated using the \bhtree{} $N$-body code \citep{barnesHierarchicalLogForcecalculation1986}.
These properties are then utilised by the Python sections of the code authored for this paper, which calculate enrichment through massive stellar winds and supernovae.
In addition to this section, code to determine initial mass functions and disk lifetimes was devised, and will be discussed in brief as well.

\subsection{General operation \& \texttt{AMUSE}}

The \amuse{} framework was used in the simulations in this paper to initialise, translate and manage the external $N$-body and stellar evolution codes \citep{zwartAstrophysicalRecipesArt2018,portegieszwartMultiphysicsSimulationsUsing2013,pelupessyAstrophysicalMultipurposeSoftware2013,portegieszwartMultiphysicsMultiscaleSoftware2009}.
While the $N$-body and stellar evolution codes in \amuse{} can be bridged and run alongside each other, the code written to calculate SLR enrichment cannot be run synchronously.
As such, the main execution loop is divided into a series of timesteps so that the enrichment routines can run in lockstep with the rest of \texttt{AMUSE}.
As closer encounters between stars are fleeting when compared to the simulation timescale, a relatively small timestep is adopted in order to sample wind injection more accurately.
We found that a total number of 1000 timesteps for the enrichment code over \SI{20}{Myr} was suitable for our simulations, the $N$-body and stellar evolution codes operate on smaller timesteps. 

The \seba{} stellar evolution code was utilised to calculate mass loss rates for main and post-main sequence evolution of stars within a star-forming region.
\seba{} is extremely fast, using an interpolated lookup table for stellar evolution rather than directly calculating the stars' properties -- such as in the case of a Henyey code like \texttt{MESA} \citep{paxtonModulesExperimentsStellar2011a,portegieszwartPopulationSynthesisHighmass1996,toonenSupernovaTypeIa2012}.
Whilst this approximation method is less accurate, \seba{} is suitable for this work due to its speed.
It is important to note that mass-loss rates due to winds can vary significantly between stellar evolution models. However, a common factor between all models is that the bulk of mass loss over the lifespan of a particularly massive star occurs after the main sequence, which is itself a comparatively short period of time ($0.5\,\si{Myr}$ to $1\,\si{Myr}$); as such most enrichment will occur over this time, and should not vary significantly between models.
\seba{} is also used to determine whether massive stars have undergone supernovae.
Dynamic evolution of the star-forming region is performed using the \bhtree{} $N$-body code.
\bhtree{} is a 2\textsuperscript{nd}-order accurate Barnes-Hut octree code designed for large numbers of particles.
Whilst pure $N$-body codes are more accurate, the extremely favourable scaling characteristics ($\mathcal{O}(N\log{N})$ vs. $\mathcal{O}(N^2)$ of pure solvers) allow hundreds of large scale simulations to be run within an acceptable length of time on a 10-core Xeon workstation \citep{barnesHierarchicalLogForcecalculation1986}.
As such, the reduction in accuracy was deemed an acceptable trade-off for this project.
Performance was further improved by using \texttt{OpenMPI} to multi-thread \bhtree{} \citep{gabriel04:_open_mpi}.
At every timestep the change in mass calculated through \seba{} was synchronised with \bhtree{} to account for mass loss affecting gravitational attraction.

After the $N$-body solver and stellar evolution simulations are evolved to the next timestep, the code written for this paper to calculate SLR enrichment is executed, as detailed in Section \ref{sec:slr-enrich}.

\subsection{Region formation}

Upon programme initialisation a star-forming region with a given number of stars ($N_\star$) and radius ($r_{c}$) is generated via a \citet{goodwinDynamicalEvolutionFractal2004} box fractal method.
The box fractal model spawns fractals by placing a ``root'' particle in the centre of a cube of side $N\rms{div}$, which spawns $N^3$ ``leaf'' cubes, which can also contain their own ``leaf'' particles.
The probability of each generation producing offspring is equivalent to $N\rms{div}^{D-3}$, fewer generations are produced with a lower fractal dimension parameter, $D$, which is included as an input parameter to the algorithm \citep{goodwinDynamicalEvolutionFractal2004}.
Lower values of $D$ lead to more substructure and a less uniform appearance, while higher values lead to a more homogeneous and spherical appearance.
For our simulations we adopt a fractal dimension of $D=2.0$, as this gives a moderate amount of substructure, which is observed in many star-forming regions \citep{2004MNRAS.348..589C,2009ApJ...696.2086S,2018MNRAS.473..849D} and in simulations \citep{2012MNRAS.420.3264G,2012MNRAS.424..377D,2013MNRAS.430..234D}. Subsequent dynamical evolution makes it difficult to ascertain the initial degree of substructure in a typical star-forming region \citep{2012MNRAS.427..637P,2020MNRAS.493.4925D}, but $D=2.0$ probably lies towards the middle of the distribution of expected values.
The initial velocities of the parent particles are drawn from a Gaussian distribution, with a small random component that scales as $N\rms{div}^{D-3}$ but reduces for each progressive generation.
After generating the positions and velocities of the stars within the star-forming region, the masses are calculated from an implementation of the Maschberger initial mass function \citep[IMF;][]{maschbergerFunctionDescribingStellar2013}.
The Maschberger IMF is described in the form of a probability density function following the formula:

\begin{equation}
  P(M_\star) \propto \frac{M_\star}{\mu}^{-\alpha} \left( 1 + \left(\frac{M_\star}{\mu}\right)^{1-\alpha} \right)^{-\beta} ,
\end{equation}

\noindent
where $P$ is the probability, $\mass_\star$ is the star mass in $\msol$, $\alpha$ is the high-mass exponent, $\beta$ is the low-mass exponent, and $\mu$ is the scale parameter.
As per Maschberger's prescription, we use values of $\alpha = 2.3$, $\beta = 1.4$ and $\mu = 0.2$.
The mass range of the Maschberger prescription is between $0.01\,\msol$ and $150\,\msol$. 
When generating a population of stars, we ensure that there is at least one high-mass star ($\mass_\star \geq 13 \, \msol$) in the resultant region and the IMF routine is repeatedly run until this condition is met \citep{nicholsonSupernovaEnrichmentPlanetary2017}.
Stochastic sampling of the IMF can result in low-mass stellar populations that (infrequently) contain massive stars \citep{parkerOstarsFormIsolation2007}. \cite{nicholsonSupernovaEnrichmentPlanetary2017} demonstrate that -- assuming a standard Cluster Mass Function of the form $N\rms{clus} \propto M\rms{clus}^{-2}$ -- these unusual low-mass populations that contain high mass stars occur as often as high-mass clusters which are almost certain to contain massive stars.   So whilst we do not simulate those clusters that do not contain massive stars (because there would be no enrichment), our enrichment distributions should be interpreted for stellar populations where some enrichment can occur.

All stars in the simulation have the same formation time; whilst there would be some margin of age difference in the stars this would introduce an additional parameter to explore, significantly increasing the required number of simulations.
Furthermore, \seba{} and by extension \amuse{} cannot be used to simulate mixed-age stellar populations within a single cluster datatype.
Finally, binarity of massive stars is not modelled, as wind-wind interactions of massive stars and how this affects dust formation and growth is not well established -- though in ideal cases there would be a significant increase in \al{} wind abundance \citep{brinkmanAluminium26MassiveBinary2023}.
We will study the effects of massive star binarity on the enrichment of low-mass stars in a future paper.

\subsubsection{Protoplanetary disks}

\begin{figure}
  \centering
  \includegraphics{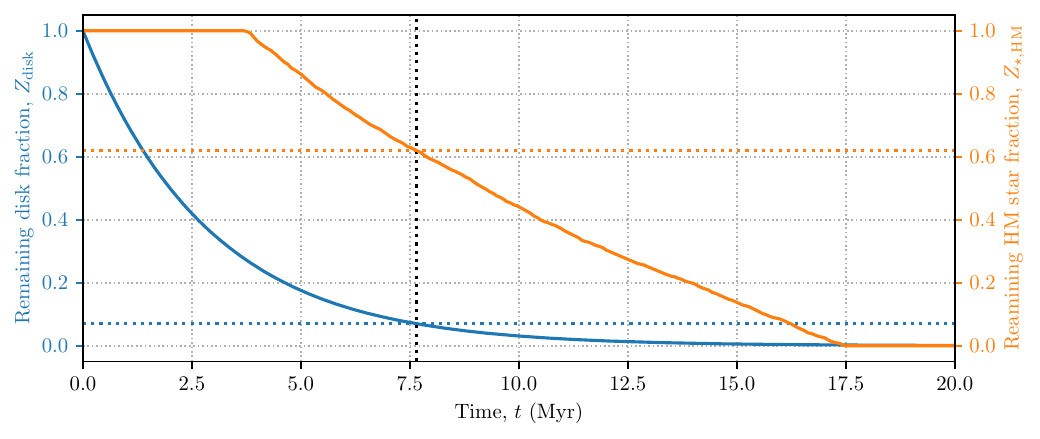}
  \caption{A comparison between the fraction of remaining disks and the fraction of high-mass stars that have not reached the end of their lives. Due to the relatively short disk lifetime most of the stellar disks have progressed to the planetesimal formation phase before most supernovae have occurred. The dotted vertical line indicates when the first stars with an initial mass under \SI{25}{\msol} undergo supernova, contributing to SLR enrichment according to \citet{limongiPresupernovaEvolutionExplosive2018}.}
  \label{fig:sne-fraction}
\end{figure}

Each low-mass star (defined as $\SI{0.1}{\msol} \leq M_\star \leq \SI{3}{M_\odot}$ throughout this paper) is assumed to form a protoplanetary disk with a radius of \SI{100}{AU}, to facilitate a direct comparison with our previous work \citep{parkerShortlivedRadioisotopeEnrichment2023,patelPhotoevaporationEnrichmentCradle2023a} -- though we note that other prescriptions for setting the disk radius as a function of stellar mass are used in the literature \citep[e.g.][]{2022MNRAS.514.2315C}. Our disks have a fixed mass dependent on the stellar mass following the formula:

\begin{equation}
  M\rms{disk} = \SI{0.1}{\mass_\star},
\end{equation}

\noindent
with a resultant dust mass of $M\rms{dust} = \SI{0.01}{\mass\rms{disk}}$.
After the dust mass is calculated we calculate the quantity of the SLR's stable isotope counterparts.
We assume a stable aluminium mass fraction of:

\begin{equation}
  \mass\rms{27Al} = 0.0085 \, \mass\rms{dust} , 
\end{equation}

\noindent
and a stable iron mass fraction of:

\begin{equation}
  \mass\rms{56Fe} = 0.1828 \, \mass\rms{dust} ,
\end{equation}

\noindent
as described in \cite{loddersSolarSystemAbundances2003}.
There is no initial quantity of the SLR isotopes \al{} and \fe{}; all enrichment is assumed to occur post-star-formation.
Disk truncation through stellar winds and ionising radiation flux are not simulated, but could be included in successive versions of this model.
The lifetime of each protoplanetary disk is pre-calculated and derived from an exponential probability density function of the form:

\begin{equation}
  p \left( x,\beta \right) = \frac{1}{\beta} e^{-x/\beta} ,
  \label{eq:disklifetime}
\end{equation}

\noindent
where $p$ is the proabability density, $\beta$ is the mean lifetime of a protoplanetary disk.
We adopt a value of $\beta = \SI{2}{Myr}$ for the mean disk lifetime, consistent with the findings in \cite{richertCircumstellarDiscLifetimes2018}.
An example of the disk population over time can be seen in Fig. \ref{fig:sne-fraction}.
Stars with masses above \SI{3}{\msol} are not assigned disks, as they are markedly less likely to form planetary systems containing rocky bodies \citep{roncoPlanetFormationIntermediatemass2024,berneFarultravioletDrivenPhotoevaporation2024}.
Final enrichment is calculated at the point where the individual disk has ``progressed'' from a disk to a protoplanetary system.
Beyond this point, the system is no longer capable of further SLR pollution, and the final enrichment value is stored.
A more advanced model could reduce the efficiency of the disk in absorbing SLRs as planetesimals begin to form, before culminating in a finished protoplanetary system, though for our work this 2-phase method should suffice.

\subsection{SLR Enrichment}
\label{sec:slr-enrich}

\begin{figure}
  \centering\includegraphics{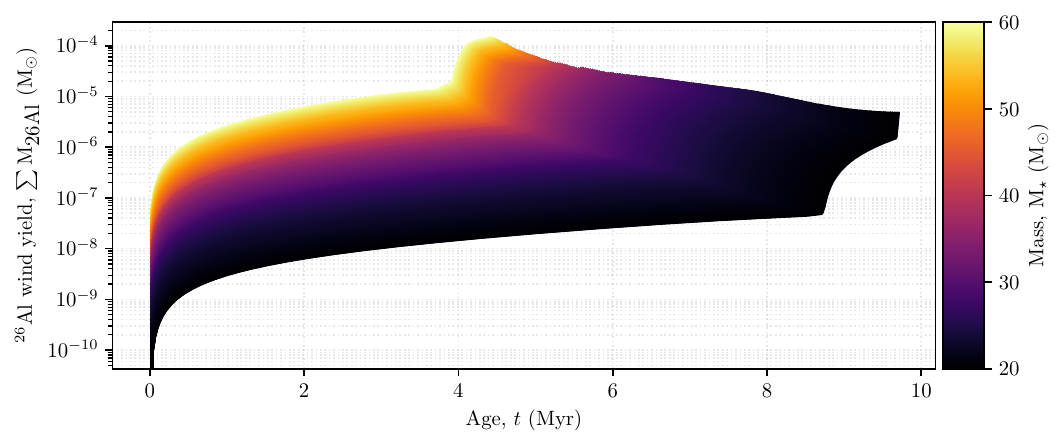}
  \caption{A comparison of cumulative yield of \al{} in stellar winds over \SI{10}{Myr} from stars with a mass range of $20-60\,\si{\msol}$. The majority of the mass loss occurs near the end of each star's life, as the star exits the main sequences and begins its Wolf-Rayet phase.}
  \label{fig:cumulativeyield}
\end{figure}

For this paper we only simulate the effect of wind and SNe-based enrichment of the proto-planetary disk. SLR enrichment from other mechanisms such as sequential star formation, spallation or molecular cloud pre-enrichment are not considered.
Estimates for wind and supernovae yields of SLRs from massive stars were derived from simulations performed in \cite{limongiPresupernovaEvolutionExplosive2018}.
For this paper we use their recommended (``R'') model with a star rotational velocity of \SI{300}{km.s^{-1}}.
There is a small difference in \al{} yield in the \SI{150}{km.s^{-1}} model and the \SI{300}{km.s^{-1}} model between \SI{13}{\msol} to \SI{20}{\msol}, though for the \SI{20}{\msol} case there is a higher yield in the \SI{150}{km.s^{-1}} model.
The \SI{0}{km.s^{-1}} model was not considered as this was unrealistic \citep[all massive stars are thought to rotate, though the exact velocity is under debate,][]{deminkRotationRatesMassive2013}, and significantly suppressed \al{} yields compared to the other two options.
The results in \cite{limongiPresupernovaEvolutionExplosive2018} provide a final total yield, using this value we calculate the fraction of SLR emitted relative to the total wind mass loss rate using \seba{} to calculate the total mass loss rate before the simulation begins.
The estimated fraction is calculated through an Akima spline interpolation of the data provided in the paper, which provides a more accurate fit compared to a cubic spline.
This fraction can then be used in conjunction with the simulation mass loss rate from \seba{} to approximate the SLR emission rate from massive stars.
Whilst directly calculating the SLR loss rate through their model would yield improved results --- as our estimation does not factor in composition changes in the wind through stellar interior processes such as dredge-up --- these issues are offset by the fact that the majority of the mass loss of early-type stars occurs after they leave the main sequence in a comparatively short period of time before supernova (see Fig. \ref{fig:cumulativeyield}).
This method also works for SLRs beyond \al{} and \fe{}, such as \atom{Cl}{36}, \atom{CA}{41} and \atom{Mn}{53}, with a minimal increase in computational time per SLR, however these were not considered for this paper. Future versions of this code will be able to analyse other SLRs, as the routines governing enrichment and decay were designed with flexibility in mind.

In the case of supernovae, the total explosive yield of SLRs is provided by \citet{limongiPresupernovaEvolutionExplosive2018}.
It is important to note that the \citet{limongiPresupernovaEvolutionExplosive2018} model assumes that stars above \SI{25}{\msol} collapse directly into a black hole, producing no supernova explosion and thus resulting in no further enrichment.
The debate on this facet of stellar evolution is still ongoing, the mass border wherein supernovae occur is still somewhat ill-defined, though contemporary research suggests that supernovae become increasingly rare above \SI{20}{\msol} \citep{ertlExplosionHeliumStars2020,ebingerPUSHingCorecollapseSupernovae2020}.
We determined that for this paper we would not consider these higher-mass supernovae -- in particular due to the rarity of higher-mass precursor stars -- though investigating the effect of higher-mass supernovae on star-forming region enrichment could be a future avenue of research.

\subsubsection{Disk enrichment from stellar winds}
\label{sec:slr-wind}

In order to determine the quantity of material swept up by the protoplanetary disk, we must calculate the geometric cross-section of the disk when interacting with the wind.
To perform this calculation we approximate the outflow as a spherical wind-blown bubble with a constant density.
From this we can calculate the sweep-up area, $\eta\rms{sweep}$, equivalent to:

\begin{equation}
  \eta\rms{sweep} = \frac{3}{4} \frac{r^2\rms{disk} \Delta r_\star}{r^3\rms{bub}} ,
\end{equation}

\noindent
where $r\rms{disk}$ is the disk radius, $\Delta r_\star$ is the distance travelled by the approaching star during a timestep and $r\rms{bub}$ is the bubble radius.
Two bubble radii are considered, a small bubble, with a radius of \SI{0.1}{pc}, and a variable bubble size with a radius equivalent to the current virial radius of the region.
These are representative of a small, local wind (the ``local'' model), and a diffuse, dispersed stellar wind throughout the region (the ``global'' model) respectively.
The total effective cross-section from wind absorption, $\eta\rms{wind}$, is given by the equation:

\begin{equation}
  \eta\rms{wind} = \eta\rms{con} \eta\rms{inj} \eta\rms{sweep} , 
\end{equation}

\noindent
where $\eta\rms{con}$ is the dust condensation efficiency and $\eta\rms{inj}$ is the disk injection efficiency.
For this paper, $\eta\rms{con} = 0.5$ \citep{lichtenbergIsotopicEnrichmentForming2016} and $\eta\rms{inj} = 0.7$ \citep{ouelletteInjectionMechanismsShortlived2009}.
These parameters can vary based on the properties of the massive star and the disk, though these represent somewhat conservative values for both \citep{matsuuraHerschelDetectsMassive2011}, with the bubble radius being the most influential free parameter.
Finally, we can calculate the mass sweep-up rate, $\dot{\Gamma}\rms{wind,SLR}$, of \al{} and \fe{} from the wind-blown bubble using the formula:

\begin{equation}
  \dot{\Gamma}\rms{wind,SLR} = \eta\rms{wind} \dot{\mass}\rms{SLR} ,
\end{equation}

\noindent
where $\dot{\mass}\rms{SLR}$ is SLR ejection rate of the massive star.

The ``global'' and ``local'' models are run concurrently, as they have no influence on the $N$-body trajectories nor stellar evolution of the stars within the star-forming region.
Additionally, radioactive decay is modelled as the simulation runtime is significantly longer than the half-life of either SLR.
Decay is calculated by determining the fraction of remaining SLRs between each timestep, based on the \al{} and \fe{} half-lifes of \SI{0.7}{Myr} and \SI{2.6}{Myr}, respectively.

\subsubsection{Disk enrichment from supernovae}
\label{sec:sne-enrich}

Supernovae enrichment is calculated directly from the supernova yield of the star.
All stars within the star-forming region are instantaneously injected with SLRs.
Whilst the outflow material from the supernova would not be travelling instantaneously, the crossing time of the outflow is on the order of \SI{100}{yr}, significantly smaller than the smallest simulation timestep of \SI{2e4}{yr}, and as such this is a reasonable abstraction.
The amount of material from the supernova deposited onto a protoplanetary disk is dependent on the cross-sectional area of the disk relative to the supernova, such that the geometric efficiency, $\eta\rms{geom}$, is:

\begin{equation}
  \eta\rms{geom} = \frac{\pi r\rms{disk}^2 \cos{\theta} }{4 \pi d^2} , 
\end{equation}

\noindent
where $d$ is the distance from the approaching star and $\theta$ is the disk inclination relative to the massive star.
We assume a constant disk inclination of $60^\circ$, such that $\cos{\theta} = 0.5$.
The total supernova SLR absorption efficiency, $\eta\rms{SNe}$, is given by the equation:

\begin{equation}
  \eta\rms{SNe} = \eta\rms{con} \eta\rms{inj} \eta\rms{geom}, 
\end{equation}

\noindent
with the same values detailed in Section \ref{sec:slr-wind}.
From this we calculate the SNe injection mass, $\Gamma\rms{SNe,SLR}$, of both SLRs with the following equation:

\begin{equation}
  \Gamma\rms{SNe,SLR} = \eta\rms{SNe} M\rms{SNe,SLR} ,
\end{equation}

\noindent
where $M\rms{SNe,SLR}$ is the SLR yield from the supernova.

\subsection{Simulation post-processing}

Enrichment through the ``global'' model and the supernovae-resultant enrichment are calculated using a post-processing method which calculates the enrichment rates at each simulation checkpoint, while enrichment through the ``local'' model is calculated at the end of each $N$-body timestep.
Both methods are very fast, with post-processing through the global model taking a few seconds per simulation on a reasonably powerful desktop PC, while the local model takes $5-10\%$ of the total computational time per simulation step on the same computer.
The main source of compute time is the calculation of distances between high-mass and low-mass stars.
Adding more SLRs for consideration would not impact performance significantly as calculating the enrichment rate is comparatively simple.
The contributions through each model are collated as part of a final post-processing step, which is performed before plots were made.

\section{Results}
\label{sec:results}

We performed repeated simulations with varying star number densities by modifying the star forming region radius and number of stars within the star-forming region.
The number of stars, $N_\star$ is varied between 100, 1000, and $10^4$ stars, in order to observe the effect of star number density ($\rho\rms{c}$) and the compound effect of multiple massive stars and supernovae.
The star-forming region radius is varied between 0.3, 1.0 and 3.0 parsecs, to provide representative examples of extremely compact, compact and dispersed star-forming regions.
This correlates to extremely dense and relatively dispersed star-forming regions, and is done in order to differentiate between effects due to region density and massive star count.
Each set of parameters is repeated 32 times, to reduce the statistical impact of systems enriched through improbable means such as multiple supernovae passes (Table \ref{tab:star-stats}).

An important factor governing the enrichment of disks within these simulations is that the rate of disk progression is significantly faster than the average rate of supernovae occurring for any given simulation.
Whilst massive stellar winds make up the bulk of \al{} enrichment, \fe{} enrichment through winds is many orders of magnitude slower, as enrichment relies more heavily on supernovae, \fe{} enrichment is curtailed.
Furthermore, as we make the same assumption as \cite{limongiPresupernovaEvolutionExplosive2018} that only stars with an initial mass below \SI{25}{\msol} produce supernovae, with stars higher than this initial mass instead directly collapsing into black holes.
As such, enrichment from supernovae does not occur until $\sim \SI{7}{Myr}$ after the start of the simulation, at a point where only $7\%$ of disks remain, as seen in Fig. \ref{fig:sne-fraction}.

\subsection{Population statistics}

\begin{table*}
  \begin{tabular}{lllllll}
  \hline
  \multicolumn{1}{c}{Set Name} & \multicolumn{1}{c}{$R\rms{c}$} & \multicolumn{1}{c}{$N_\star$} & \multicolumn{1}{c}{$N_{\star,\text{HM}}$} & \multicolumn{1}{c}{$N\rms{SNe}$} & \multicolumn{1}{c}{$Z\rms{local,enrich}$} & \multicolumn{1}{c}{$Z\rms{26Al,0.1SS}$}\\ \hline
  \texttt{pt-0.3-100-fr} & 0.3 & 100 & $1.125\pm0.074$ & $0.781\pm0.098$ & $0.223\pm0.022$ & $0.063 \pm 0.021$ \\
  \texttt{pt-0.3-1000-fr} & 0.3 & 1000 & $2.719\pm0.278$ & $1.500\pm0.201$ & $0.426\pm0.027$ & $0.046 \pm 0.012$ \\
  \texttt{pt-0.3-10000-fr} & 0.3 & 10000 & $20.031\pm0.667$ & $13.531\pm0.582$ & $0.877\pm0.003$ & $0.502 \pm 0.026$ \\
  \texttt{pt-1.0-100-fr} & 1.0 & 100 & $1.062\pm0.043$ & $0.844\pm0.079$ & $0.015\pm0.003$ & $0.003 \pm 0.002$ \\
  \texttt{pt-1.0-1000-fr} & 1.0 & 1000 & $2.188\pm0.213$ & $1.375\pm0.140$ & $0.023\pm0.003$ & $0.003 \pm 0.001$ \\
  \texttt{pt-1.0-10000-fr} & 1.0 & 10000 & $19.344\pm0.823$ & $12.531\pm0.592$ & $0.244\pm0.007$ & $0.024 \pm 0.003$ \\
  \texttt{pt-3.0-100-fr} & 3.0 & 100 & $1.094\pm0.052$ & $0.844\pm0.091$ & $0.001\pm0.001$ & $0.001 \pm 0.001$ \\
  \texttt{pt-3.0-1000-fr} & 3.0 & 1000 & $2.594\pm0.245$ & $1.906\pm0.203$ & $0.001\pm0.000$ & $0.000 \pm 0.000$ \\
  \texttt{pt-3.0-10000-fr} & 3.0 & 10000 & $21.000\pm0.779$ & $14.438\pm0.654$ & $0.015\pm0.001$ & $0.001 \pm 0.000$ \\   
  \end{tabular}
  \caption{A table showing the simulation parameters as well as a number of key parameters, the mean high mass star count in the simulation $N_{\star\text{,HM}}$, the mean number of supernovae by the end of the simulation $N\rms{SNe}$, the fraction of disks enriched by the local model, $Z\rms{local,enrich}$, and the fraction of disks undergoing \al{} enrichment approaching Solar System estimates ($0.1\times$ Solar value), $Z\rms{26Al,0.1SS}$. It is important to note that for 100-star simulations there is a significantly higher fraction of massive stars, due to the requirement of a high-mass star in each simulation.}
  \label{tab:star-stats}
\end{table*}

Table \ref{tab:star-stats} contains detailed statistics of each simulation set, in particular the number of high-mass stars ($N_{\star\text{,HM}}$), simulation supernovae count ($N\rms{SNe}$), fraction of disks undergoing enrichment, ($Z\rms{local,enrich}$) and the fraction of disks that underwent enough enrichment for significant degassing to occur (defined as $0.1\times$ Solar System enrichment, $Z\rms{26Al,0.1SS}$). Enrichment calculations are based on the local model.

\begin{figure}
  \centering
  \includegraphics{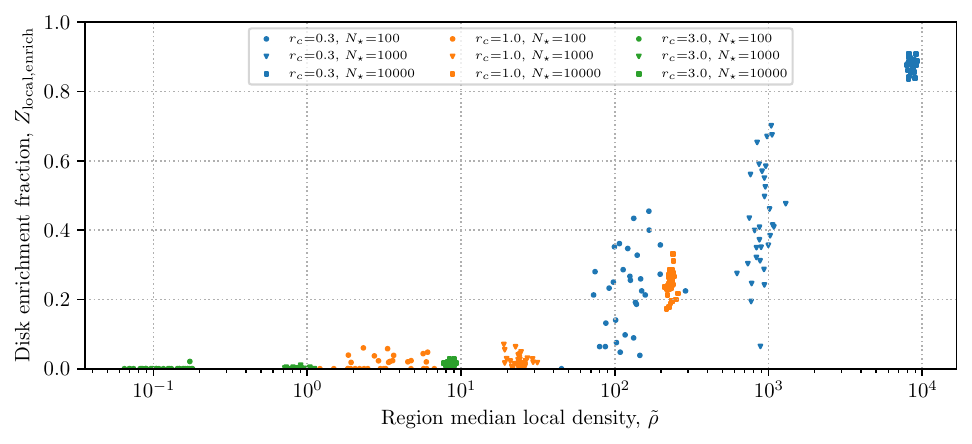}
  \caption{A plot of all simulations showing the fraction of all stellar disks that underwent any amount of enrichment from the local wind model versus initial median local density. We observe a clear dependence on region density, which is more important than the number of stars in the region, despite simulations with a higher number of stars inherently having more massive stars.}
  \label{fig:zero-yield-counts}
\end{figure}

To calculate the density of a star forming region in these simulations we use the initial median local density \citep{parkerDynamicsStructureBreaking2014,parkerFarExtremeUltraviolet2021}.
This value is calculated by determining the median value of the stellar density of a sphere containing the 10 nearest stars to each star in the star forming region at the start of the simulation.
Figure \ref{fig:zero-yield-counts} compares the fraction of all disks that underwent any form of enrichment under the local model versus the median local density.
This non-zero enrichment fraction is used to show how likely close passes to massive stars are for disks.
We find that the amount of disks undergoing enrichment for any given simulation is highly dependent on the initial median local density of the star forming region.
This suggests that a compact, low population star forming region with a single massive star is potentially ideal for SLR enrichment.
Such region masses would be significantly more likely to occur in nature than higher-mass star forming regions, and the number of low-mass regions containing a single massive star is similar to the number of higher mass regions that always contain massive stars \citep{nicholsonSupernovaEnrichmentPlanetary2017}.
Furthermore, the lower massive star count would result in less disk disruption due to ionising photon flux and supernovae shocks \citep{2019MNRAS.485.4893N,patelPhotoevaporationEnrichmentCradle2023a}.
Even in the case of extremely large numbers of stars, region radius and density appear to be more influential than star counts in keeping massive stellar winds distant from disks.
It should be noted that the 100-star systems have a higher fraction of massive stars to low mass stars, due to the minimum simulation requirement of one massive star.
The percentage of massive stars in simulations with $100$ stars is $\sim 1.1\%$ compared to $\sim 0.25\%$ of the more populous simulations.
This can potentially overestimate the amount of enrichment in a particular system; however, simulating systems without an SLR source would have been redundant.

\subsection{Comparison between models}

\begin{figure}
  \centering
  \includegraphics{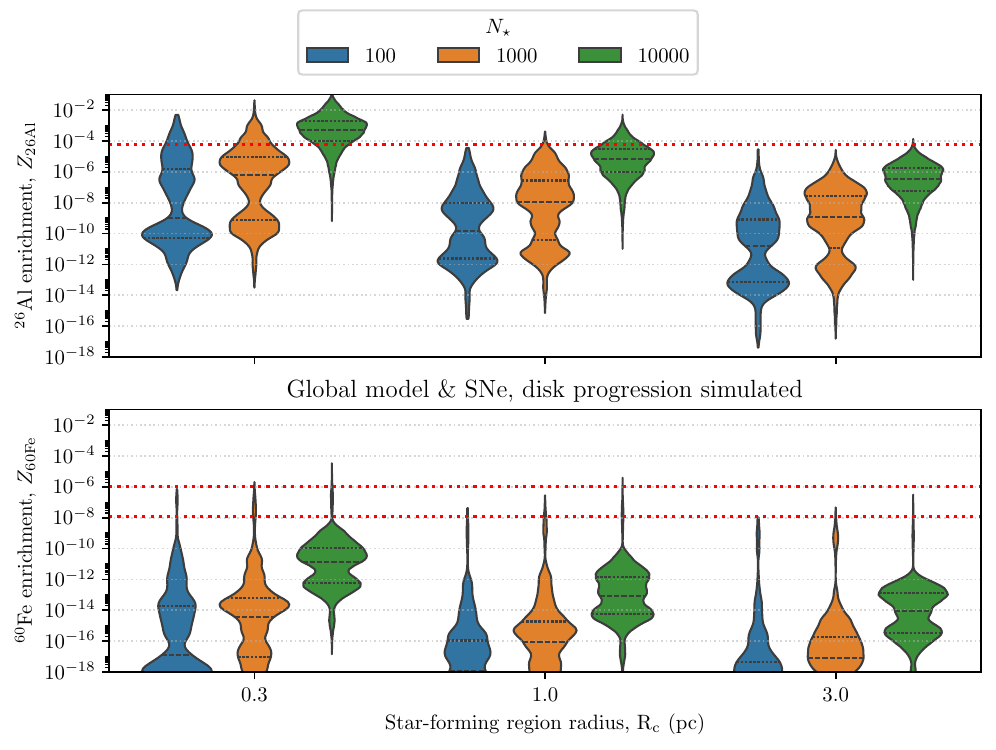}
  \caption{A series of violin plots detailing \al{} and \fe{} enrichment. Solar System enrichment is shown as a red dashed line, a value of $Z\rms{26Al,SS} \approx 5\times10^{-5}$ is used for the \al{} graphs \citep{thraneExtremelyBriefFormation2006}, while a low value of $Z\rms{60Fe,SS} = 10^{-8}$ and a high value of $Z\rms{60Fe,SS} = 10^{-6}$ are used for \fe{} plots \citep{tangAbundanceDistributionOrigin2012,mishraAbundance60FeInferred2016}. The enrichment values are calculated at the point where the disk has ``progressed'' or at the end of the simulation, whichever is shortest. While there is a higher amount of \al{} enrichment; the primarily supernovae-driven \fe{} enrichment is suppressed as most disks may not undergo supernovae interaction.}
  \label{fig:violin-global-decay}
\end{figure}

Fig. \ref{fig:violin-global-decay}
shows enrichment values
calculated at the end of the disk's lifetime.
Wind enrichment is influential, as can be seen in the \al{} results, as \fe{} enrichment is extremely minimal (though not zero) through stellar winds.
We see that for low radius star-forming regions there are a significant number of stars with \al{} enrichment greater than the Solar System estimate, however for regions with a higher radius we see that this drops off rapidly, with only the simulation set where $r\rms{c} = 1\,\si{pc}$ and $N_\star = 10^4$ exceeding the Solar System average.

In the case of \fe{} enrichment there is an even greater discrepancy between the two models, as the \fe{} yield from SNe is multiple orders of magnitude greater than the total wind yield from a massive star, \fe{} enrichment relies almost entirely on supernovae.
As disk population is reduced when supernovae begin to occur (see Fig. \ref{fig:sne-fraction}) we find that there is a vanishingly small population of stars that have Solar System-like levels of \fe{} enrichment.
The longer half-life of \fe{} does not contribute much to increasing the final enrichment of disks, whereas for \al{} it was extremely significant.
As we discussed in our previous paper \citep{eatsonDevolatilizationExtrasolarPlanetesimals2024}, \fe{} enrichment is only important for the evolution and desiccation of planetesimals in the case of extremely high levels of enrichment ($Z_\text{60Fe} \gtrapprox 10^{-2}$ or $10^{4} \times Z\rms{60Fe,SS}$). Our results in Fig. \ref{fig:violin-global-decay} show that these levels of enrichment would be almost impossible, and nothing even close was observed in our suite of simulations.

We can infer from these results that disk lifetimes are highly influential in the calculation of SLR enrichment.
SLRs that make up a significant fraction of the stellar wind, such as \al{}, are continuously distributed over the lifespan of the massive stars within the star-forming region.
However, the shorter half-life of \al{} results in significantly skewed results if the measurement of final \al{} enrichment occurs at the end of the simulation.
Furthermore, disks that survive long enough that a local star enters the Wolf-Rayet phase and goes supernova will see a significantly increased \al{} enrichment, as well as some \fe{} enrichment.

\begin{figure}
  \centering
  \includegraphics{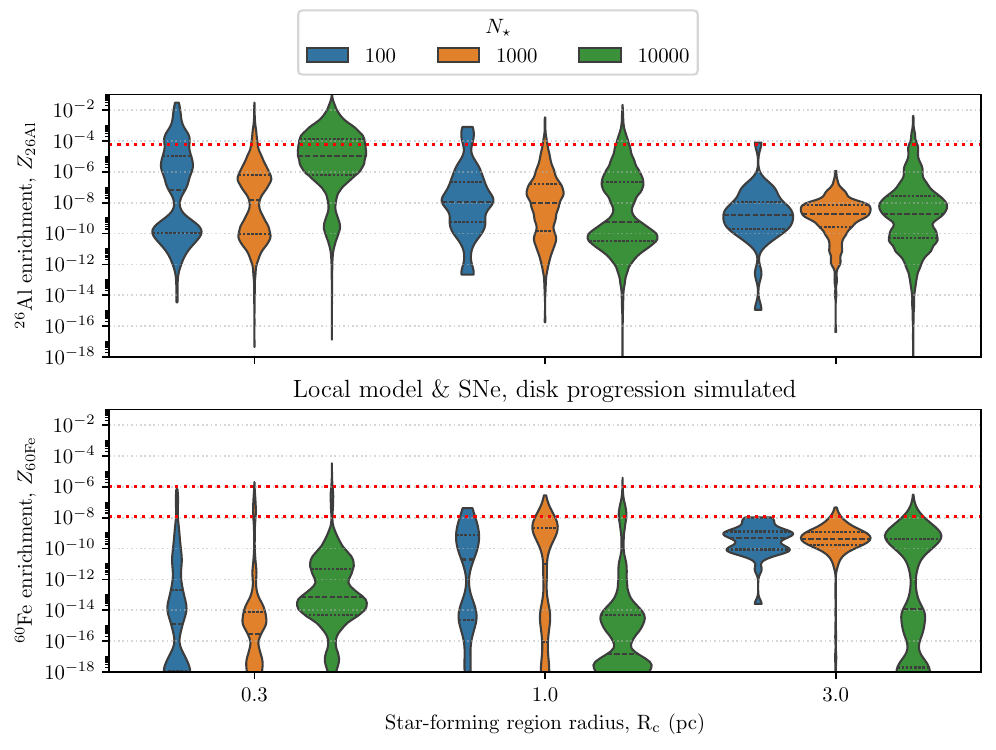}
  \caption{A series of violin plots showing \al{} and \fe{} enrichment, enrichment is calculated through the local model while disk progression is simulated in a similar manner to Fig. \ref{fig:violin-global-decay}. Solar System enrichment is shown as a red dashed line, a value of $Z\rms{26Al,SS} \approx 5\times10^{-5}$ is used for the \al{} graphs \citep{thraneExtremelyBriefFormation2006}, while a low value of $Z\rms{60Fe,SS} = 10^{-8}$ and a high value of $Z\rms{60Fe,SS} = 10^{-6}$ are used for \fe{} plots \citep{tangAbundanceDistributionOrigin2012,mishraAbundance60FeInferred2016}. Compared to the global model SLR enrichment is significantly elevated; however many simulations undergo no enrichment whatsoever with this model (see Table \ref{tab:star-stats}).}
  \label{fig:violin-local-decay}
\end{figure}

\begin{figure}
  \centering
  \includegraphics{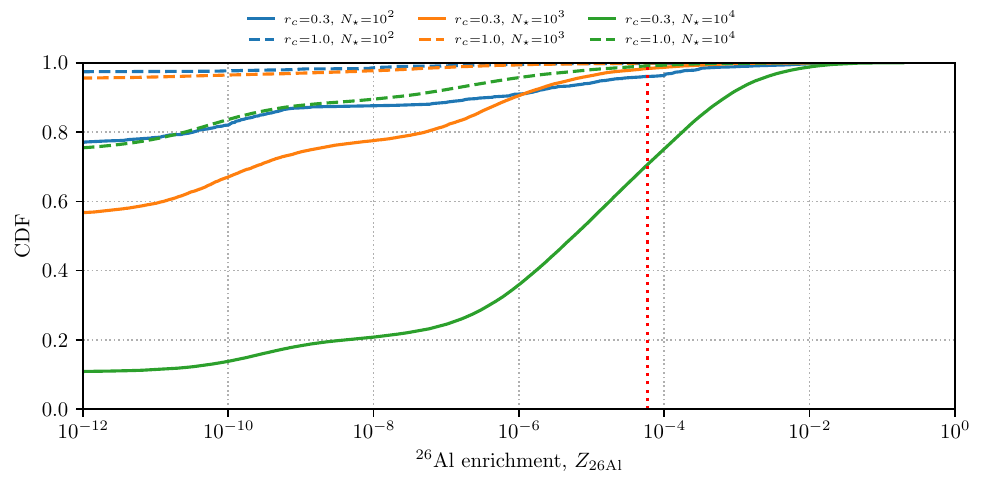}
  \caption{A cumulative distribution function (CDF) of final \al{} enrichment fraction ($Z\rms{Al}$) for all disks grouped by subset with disk progression simulated. A significant population of disks have a higher enrichment fraction than the Solar System upon entering the planetesimal formation phase in the cases of systems with extremely large numbers of stars.}
  \label{fig:cdf-al26-localsne}
\end{figure}

Fig. \ref{fig:violin-local-decay} shows the results for simulations where the ``local'' model is used instead, we observe that a greater number of disks become highly \al{} enriched, and in some cases can be enriched more than two orders of magnitude higher than the Solar System baseline. 
However, it should be noted that with the ``local'' model a disk system can have zero enrichment, which would not be represented in the violin plot.
Enrichment fractions of disks for each simulation set are shown in Table \ref{tab:star-stats}.
This enrichment fraction is shown in more detail in Fig. \ref{fig:cdf-al26-localsne}, where we can clearly see a small fraction of disks in compact star-forming regions being enriched to above-Solar System amounts, even in the case of regions with a low overall population of stars.
Above-Solar System enrichment can occur in star-forming regions with a higher initial radius, though this is significantly less likely, and is similarly independent of the total number of stars in the region.
The total number of systems undergoing \fe{} enrichment is similarly changed, as the SLR flux onto a disk is significantly increased in the local model when a disk is proximal to a massive star.
However, this effect does not drastically increase the number of systems with Solar System-like levels of \fe{}, which is still dominated by supernovae rather than stellar winds, and as such the impact of accounting for disk lifetimes has a more marked effect on enrichment than the wind model.

From previous work in \citet{lichtenbergWaterBudgetDichotomy2019} and \citet{eatsonDevolatilizationExtrasolarPlanetesimals2024} it is found that \al{} is highly important in the process of planetesimal desiccation, with enrichment levels greater than $0.1\times Z\rms{26Al,SS}$ causing a reduction in final water retention fraction of larger early planetesimals.
Table \ref{tab:star-stats} shows that only a small fraction of disks gain this level of enrichment, requiring compact, dense star-forming regions.
The main outlier is the simulation set of regions with $10^4$ stars and radii of \SI{0.3}{pc}; these super-dense star-forming regions frequently produce highly enriched disks, though occurrence of such massive star-forming regions is rare, and also the large number of supernovae that occur would result in significant disk disruption \citep{nicholsonSupernovaEnrichmentPlanetary2017}.
It is also important to note that at a radius of $3\,\text{pc}$ sufficient enrichment becomes improbable, happening to $<0.1\%$ of disks, even in the case of extremely large star counts and supernovae counts, as these values are inclusive of supernovae deposition.
\fe{} enrichment is markedly less influential to planetesimal evolution, due to the radioisotopes lower decay rate, decay energy and disk mass fraction.
In our previous paper we found that extreme levels of \fe{} enrichment ($>10^3 \times Z\rms{60Fe,SS}$) are required to produce the level of heating that results in dessication \citep{eatsonDevolatilizationExtrasolarPlanetesimals2024}.
We find in this paper that Solar System levels of enrichment are unlikely, let alone enrichment to $10^3$ times Solar System amounts.
This further rules out \fe{} as an influential SLR for planetesimal heating, and the lack of \fe{} enrichment to Solar System amounts in our simulations suggests the presence of alternative mechanisms of \fe{} enrichment beyond supernovae --- such as AGB stars \citep{parkerIsotopicEnrichmentPlanetary2023}.
Finally, comparing this result to previous work such as \citet{parkerShortlivedRadioisotopeEnrichment2023} we find that \al{} contribution from stellar winds is very significant, and that stellar density is very important.
As we vary both the star forming region and total population we find that more compact star forming regions produce greater levels of both \al{} and \fe{} enrichment, and radius is more influential on enrichment than population, especially in the case of the local wind bubble model.

\section{Discussion}
\label{sec:discussion}

Comparing the disks generated in this simulation with the Solar protoplanetary disk, we find that initial Solar System abundances for both \al{} and \fe{} do occur, but are typically less common.
This suggests that the Solar System is on the upper-end of SLR enrichment, except in the case of very dense, populous ($N_\star>10000$) star-forming regions. 
These results also suggest that the probability of ocean worlds is relatively high, as radioisotopic heating would be insufficient to cause significant devolatilization in planetesimals, resulting in a significantly higher water budget for a ``typical'' nascent protoplanet \citep{lichtenbergWaterBudgetDichotomy2019,eatsonDevolatilizationExtrasolarPlanetesimals2024}.
An expanded, more quantitative study would be necessary to estimate the population of ocean worlds in more detail. In recent years the physical cause of the density dichotomy between super-Earths and sub-Neptune exoplanets has come under closer scrutiny by formation models that involve water enrichment during the disk phase \citep[e.g.,][]{2024arXiv240401967V,2024NatAs...8..463B} instead of solely relying on post-disk atmospheric escape. Hence, a population synthesis or statistical approach involving \al{} desiccation with more modern formation models \citep[e.g.,][]{2022NatAs...6.1296K} appears a natural extension for future work to interpret the growing data from the low-mass exoplanet census \citep{2024arXiv240504057L}.

Star count does not particularly increase the maximum \al{} enrichment, though can result in more disks being significantly enriched; instead we find that star-forming region mass density significantly affects the final disk enrichment.
This is the case even with the ``global'' wind model, as the rate of SLR deposition is significantly curtailed as the bubble radius is larger.
In the more common case of compact, low mass star-forming regions, we find that around $5\%$ of disks have undergone levels of enrichment on the order of our Solar System, enough for \al{} decay heating to have a pronounced effect on planetesimal evolution, particularly affecting retained water fraction.
Whilst massive stellar winds and supernovae are a prominent source of SLRs, the adverse affects of UV photoevaporation and supernovae shocks can disrupt disks before planetesimals form \citep{patelPhotoevaporationEnrichmentCradle2023a}, making (typically low-mass) star-forming regions with fewer massive stars more conducive to planet formation.
As such, the combination of high levels of SLRs in lower population star-forming regions would result in more water poor rocky planets.
Incorporating disk progression and measuring the final enrichment from the moment the disk has decayed is a more accurate method of determining enrichment, and in the case of \al{} with its significantly shorter half-life, this can significantly impact results compared to an observation at an arbitrary time or the end of the simulation.
In the case of \fe{} however we find that supernova-based enrichment is severely inhibited by the disk lifetime typically being significantly shorter than the lifetime of high-mass stars.
We find that by the time supernovae begin to occur only a quarter of protoplanetary disks remain (Fig. \ref{fig:sne-fraction}), this is also seen in our results where disk ``decay'' skews the \fe{} yields down more than any contributing factor.
This is further inhibited when factoring in high-mass stars failing to undergo supernovae.
These results are corroborated by the findings of \cite{williamsLikelihoodSupernovaEnrichment2007} and \cite{gounelleOriginShortlivedRadionuclides2008}, whose results suggest a similar unlikelihood of Solar System level enrichment, and only in large clusters with very high mass stars.
The ``R'' model described in \cite{limongiPresupernovaEvolutionExplosive2018} assumes that direct black hole collapse occurs when a star with an initial mass greater than \SI{25}{\msol} dies, resulting in zero SLR yield.
At the point where these contributing supernovae begin to occur, less than 10\% of all disks remain, resulting in further inhibited \fe{} enrichment.
Whilst our model does not take into account mixed-age stellar populations, we do not believe that this would significantly impact our results, and would be offset by other unsimulated interactions, such as evaporation of disks from UV flux from massive stars and supernovae shocks \citep{patelPhotoevaporationEnrichmentCradle2023a}.
Furthermore, the ``low'' ($Z\rms{60Fe/56Fe,SS} \approx 10^{-8}$) \fe{} solar system estimate from \cite{tangAbundanceDistributionOrigin2012} can be explained through processes such as residual galactic chemical evolution; this mechanism does not explain the higher enrichment levels determined by other studies, such as \cite{mishraAbundance60FeInferred2016} and \cite{cookIronNickelIsotopes2021}. It appears unlikely that these higher estimates can be explained purely through supernovae. Alternative \fe{} injection mechanisms such as interloping AGB stars could prove a suitable explanation for this discrepancy \citep{trigo-rodriguezRoleMassiveAGB2009,karakasStellarYieldsMetalRich2016,parkerIsotopicEnrichmentPlanetary2023}.

\section{Conclusion}
\label{sec:conclusion}

In this paper we performed a series of simulations to determine typical enrichment rates of star-forming regions for specific densities and stellar populations.
Results differ from previous work \citep{parkerShortlivedRadioisotopeEnrichment2023,lichtenbergIsotopicEnrichmentForming2016},
as whilst we observe a similar amount of \al{} enrichment, the local wind model produces a higher number of disks undergoing \al{} enrichment significant enough to influence planetary evolution.
Much of the changes in enrichment compared to previous work are due to the length of time required for supernovae to occur, which is much longer than the average time a disk needs to progress.
Stratification of enrichment from varying simulation parameters is very clear, with enrichment increasing with star-forming region mass density.
We also find that enrichment method and disk model are particularly influential on the final SLR enrichment amount.
\al{} enrichment is primarily through wind deposition --- as this is relatively consistent for either model assuming the system is sufficiently populous or dense ---  when compared to supernovae deposition.
\fe{} enrichment, however, is almost entirely dependent on supernovae deposition, and as such enrichment to Solar System levels is unlikely, and enrichment to levels where \fe{} is the dominant heating source of planetesimals is almost non-existent.

In the immediate future, including additional SLR sources such as from ``interloping'' AGB stars would provide an interesting comparison.
As \fe{} is formed in the core of more massive AGB stars and can be efficiently dredged up and deposited via the stellar wind -- this could result in another avenue of \fe{} enrichment in lieu of the relatively ineffective massive stellar wind mechanism \citep{parkerIsotopicEnrichmentPlanetary2023}.

\section*{Acknowledgments}

\begin{acknowledgments}
JWE and RJP acknowledge support from the Royal Society in the form of a Dorothy Hodgkin fellowship, and associated enhancement awards. TL acknowledges support by the Branco Weiss Foundation, the Alfred P. Sloan Foundation (AEThER project, G202114194), and NASA's Nexus for Exoplanet System Science research coordination network (Alien Earths project, 80NSSC21K0593).
\end{acknowledgments}

\bibliography{references,TL_refs1,TL_refs2}{}
\bibliographystyle{aasjournal}

\end{document}